\documentstyle[pre,twocolumn,aps,epsf,floats]{revtex}

\begin{document} 
\draft

\title{Density fluctuations and phase transition in the Nagel-Schreckenberg
traffic flow model}

\author{S. L\"ubeck, M.~Schreckenberg, and K.~D. Usadel}

\address{
Theoretische Physik, 
Gerhard-Mercator-Universit\"at Duisburg,\\ 
Lotharstr. 1, 47048 Duisburg, Germany \\}

\author{\vskip -2\baselineskip\small(Received 15 March 1997; revised
manuscript received 23 July 1997)\break}
\author{\parbox{14cm}{\small
\quad We consider the transition of the Nagel-Schreckenberg 
traffic flow model from the free flow regime to the 
jammed regime.
We examine the inhomogeneous character of the system
by introducing a new method of analysis which is based 
on the local density distribution.
We investigated the characteristic fluctuations in the steady state
and present the phase diagram of the system. 
\hfill\break
\leftline{PACS numbers: 89.40.+k,05.40.+j,05.60.+w}}}
\address{\vskip -0.5\baselineskip}

%
%
%
%
%
%
%

\maketitle

\setcounter{page}{1}
\markright{\rm
Phys. Rev.~E {\bf 56}, (1997)
, (to be published).
[accepted for publication]  }
\thispagestyle{myheadings}
\pagestyle{myheadings}

\section{Introduction}

Over the past few years much attention has been devoted to 
the study of traffic flow. 
Since the seminal work of Lighthill and Whitham 
in the middle of the 50's \cite{LIGHT_1} many attempts have been
made to construct more and more sophisticated models which incorporate
various phenomena occurring in real 
traffic (for an overview see \cite{WOLF}).
Recently, a new class of models, based on the idea of cellular
automata, has been proven to describe traffic dynamics
in a very efficient way \cite{NASCH_1}.
Especially the transition from free flow to jammed
traffic with increasing car density could be investigated
very accurately.
Nevertheless, besides various indications \cite{CSANYI_1}, 
no unique description
for a dynamical transition could be found.
Furthermore, no satisfying order parameter could be defined
so far.
In this article we introduce a new method of analysis
which allows us to identify the different phases of the system
and to describe the phase transition in detail, i.e., defining an
order parameter, considering the fluctuations which drive the transition,
and determining the phase diagram.


We consider a one-dimensional cellular automaton of
linear size $L$ and $N$ particles.
Each particle is associated the integer values 
$v_i\in\{0,1,2,...,v_{max}\}$ and $d_i\in\{0,1,2,3,...\}$,
representing the velocity and the distance to the next
forward particle \cite{NASCH_1}.
For each particle, the following four
update steps representing the acceleration, the slowing down, the noise,
and the motion of the particles are done in parallel:
(1) if $v_i < d_i$ then $v_i \to \mbox{Min}\{v_i+1, v_{max}\}$,
(2) if $v_i > d_i$ then $v_i \to d_i$,
(3) with probability $P\/$ $v_i \to \mbox{Max}\{v_i-1, 0\}$, 
and
(4) $r_i \to r_i+v_i$, 
where $r_i$ denotes the position of the $i$-th particle.

\section{Simulations and Results}

\begin{figure}
 \begin{minipage}[b]{8.6cm}
 \epsfxsize=8.6cm
 \epsfysize=7.5cm
 \epsffile{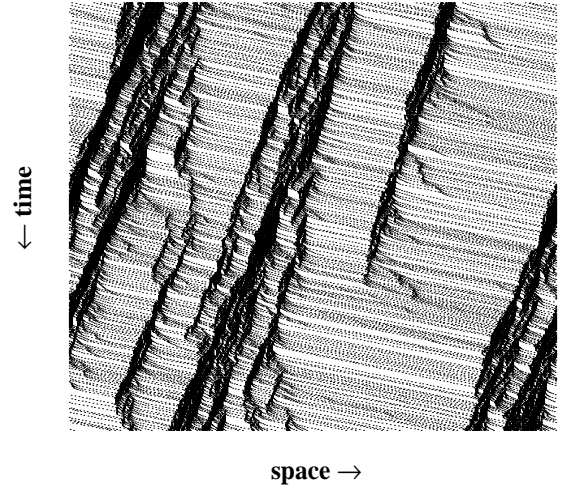} 
 \caption{Space-time plot for $v_{max}=5$, $P=\frac{1}{2}$, and $\rho_g>\rho_c$.
          Note the separation of the system in high and low density
	  regions.
 \label{space_time_dia}} 
 \end{minipage}
\end{figure}

Figure~\ref{space_time_dia} shows a space-time plot of the system.
Each dot corresponds to a particle at a given time step.
The global density $\rho_g=N/L$ exceeds the critical density and 
jams occur.
Traffic jams are characterized by a high local density of the
particles and by a backward movement of shock waves \cite{LIGHT_1}.
One can see from Fig.\ref{space_time_dia} that in the jammed regime 
the system is inhomogeneous, 
i.e., traffic jams with a high local density and free flow regions with
a low local density coexist.
In order to investigate this transition one has to take this
inhomogeneity into account.

Traditionally one determines the so-called fundamental diagram, 
i.e.,~the diagram of the flow vs the density.
The global flow is given by, 
$\Phi \; = \; \rho_g \, \langle v \rangle$,
where $\langle v \rangle$ denotes the averaged velocity of the 
particles.
Due to the stochastic behavior of the dynamics
for $0<P<1$ the system behaves independently of the initial 
conditions after a certain transient regime.
In this limit one can interpret $\langle ... \rangle$ 
as a time or ensemble average.
For $P=0$ and $P=1$ the dynamics is deterministic and the 
behavior depends strongly on the initial conditions.
In any case, this non-local measurements are not sensitive to the inhomogeneous
character of the system, i.e.,~the information
about the two different coexisting phases is lost.
In the following we apply
a new method of analysis which is based on the measurement of 
the local density distribution $p(\rho)$.
The local density $\rho$ is measured on a section of the 
system of size $\delta$ according to
\begin{equation}
\rho \; = \; \frac{1}{\rho_g \delta} \, \sum_{i=1}^{N} \,
\theta(\delta -r_i).  
\label{eq:density_dist}
\end{equation}
Of course we have checked that the main results are not
affected by the value of $\delta$, provided that delta
is significantly smaller than the system size $L$ in 
order to measure the local properties.
In order to reflect the behavior of the low density
regime $\delta$ should be significantly larger
than a certain length scale $\lambda_0$ which corresponds 
to the characteristic length scale of the density 
fluctuations in the free flow phase (see below).
For any parameter set $\{v_{max},P\}$ 
the local density $\rho$ fluctuates around the value of the
global density $\rho_g$ and the probability distribution 
of the local density $p(\rho)$ contains all informations
needed to describe the transition.

\begin{figure}
 \begin{minipage}[t]{8.6cm}
 \epsfxsize=8.6cm
 \epsfysize=7.5cm
 \epsffile{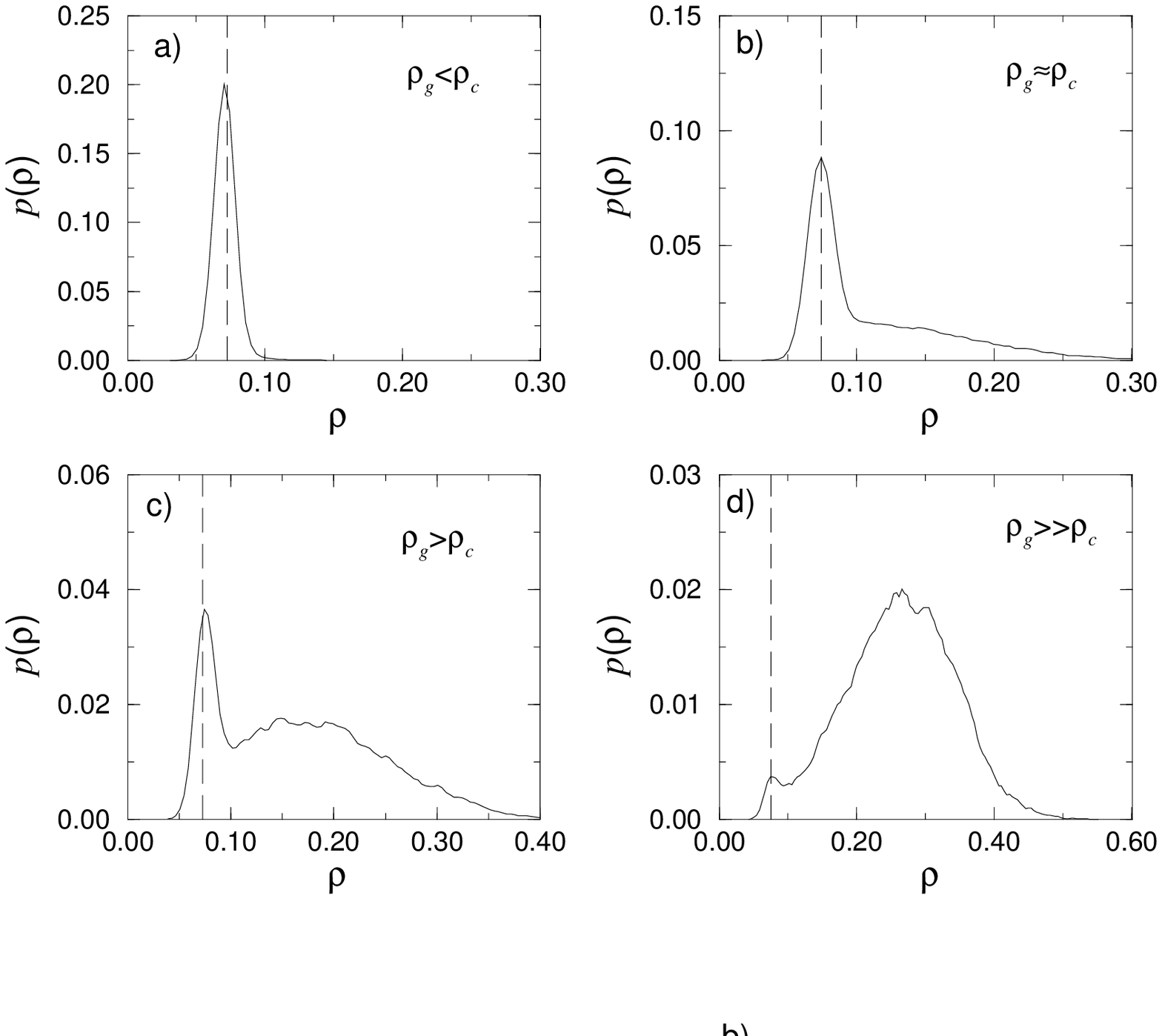} 
 \caption{The local density distribution $p(\rho)$ for various
	  values of the global density, $v_{max}=5$, $P=\frac{1}{2}$
	  and $\delta=256$. 
	  The dashed line corresponds to the characteristic density 
	  of the free flow phase.
 \label{density_dist}} 
 \end{minipage}
\end{figure}

The local density distribution $p(\rho)$ is plotted 
for various values of the global 
density $\rho_g$ in Fig.~\ref{density_dist}.
In the case of small values of $\rho_g$, see Fig.~\ref{density_dist}a, 
the particles can
be considered as independent (see below) and the 
local density distribution is simply Gaussian with
the mean values $\rho_g$ and a width which scales 
with $\sqrt{\delta}$. 
Increasing the global density, jams occur
and the distribution displays two different 
peaks (Fig.~\ref{density_dist}c).
The first peak corresponds to the density of free particles and
in the phase coexistence regime the position of this peak
does not depend on the global density (see the dashed lines
in Fig.~\ref{density_dist}).
The second peak is located at larger densities and characterizes
the jammed phase.
With increasing density the second peak occurs in the vicinity
of the critical density $\rho_c$ (Fig.~\ref{density_dist}b)
and grows further (Fig.~\ref{density_dist}c) until it dominates
the distribution in the sense that the first peak 
disappears (Fig.~\ref{density_dist}d).
The two peak structure of the local density distribution clearly 
reflects the coexistence of the free flow and 
jammed phase above the critical value $\rho_c$.
In the following we show that the behavior of the
peaks leads to a determination of $\rho_c$.

\begin{figure}
 \begin{minipage}[t]{8.6cm}
 \epsfxsize=8.6cm
 \epsfysize=7.5cm
 \epsffile{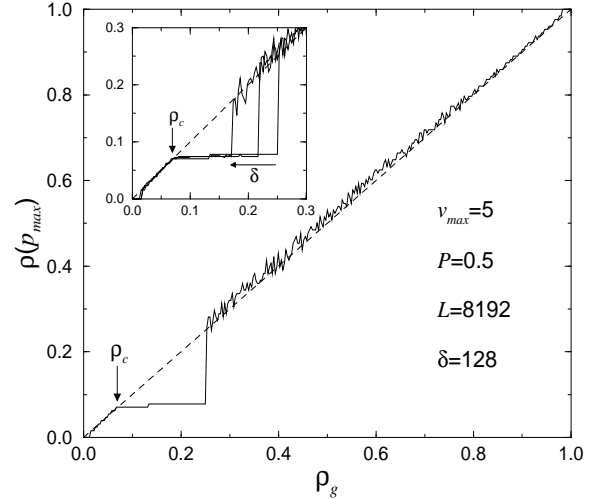} 
 \caption{The maximum of the local density distribution as a function
	  of the global density $\rho_g$. The inset shows that the
	  transition point $\rho_c$ does not depend on the value
	  of the parameter $\delta$.         
 \label{decomposition}}
 \end{minipage}
\end{figure}

One expects that in a homogeneous system
the local density distribution displays one peak and 
is symmetric around the global density, i.e., 
$\rho(p_{max})=\rho_g$.
This cannot be the case in an inhomogeneous system where
the local density distribution displays two
peaks corresponding to two different coexisting phases.
In Fig.~\ref{decomposition} we plot the position of the maximum
of the local density distribution $\rho(p_{max})$ as a
function of the global density $\rho_g$.
One clearly sees the transition point $\rho_c$ where
the position of the maximum becomes independent of the
global density.
The inset of Fig.~\ref{decomposition} shows that the
determination of the transition point does not
depend on the special value of $\delta$.
Only the point where the second peak exceeds the first peak
depends on the measurement parameter $\delta$.
With increasing $\delta$ this point tends to smaller values 
of $\rho_g$ because with increasing $\delta$ the measurement
starts to average over the two different phases.
From these measurements we conclude that the 
phase transition of the Nagel-Schreckenberg model
is a transition from a homogeneous regime (free flow phase)
to an inhomogeneous regime which is characterized by a 
coexistence of two phases (free flow traffic and jammed traffic).

\begin{figure}
 \begin{minipage}[t]{8.6cm}
 \epsfxsize=8.6cm
 \epsfysize=7.5cm
 \epsffile{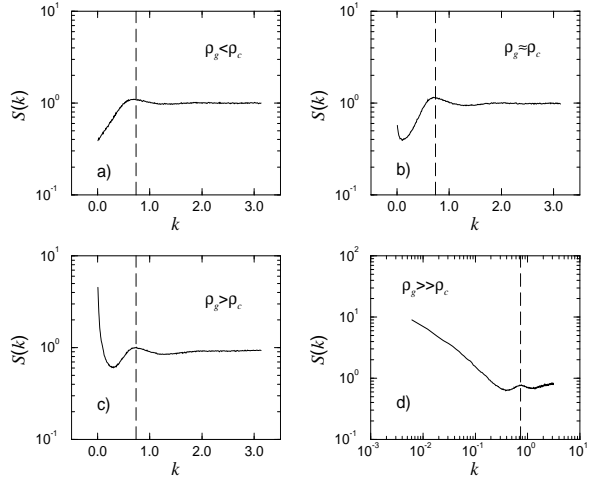} 
 \caption{The structure factor $S(k)$ for $P=\frac{1}{2}$, $v_{max}=5$
	  and for various values of the
	  density $\rho$. The dashed line marks the characteristic
	  wavelength $\lambda_0$ of the free flow phase.
 \label{struc_5}} 
 \end{minipage}
\end{figure}

In order to describe the spatial decomposition of the
coexisting phases we measured the steady state structure factor \cite{SCHMITT_1}
\begin{equation}
S(k) \; = \; \frac{1}{L} \left \langle \left | \, \sum_{r=1}^{L} \,
\eta(r) \, e^{i k r} \right|^2 \right \rangle,
\label{eq:structure_factor}
\end{equation}
where $\eta(r)=1$ if the lattice site $r$ is occupied and
$\eta(r)=0$ otherwise.
In Fig.~\ref{struc_5} we plot the structure factor $S(k)$
for the same values of the global density as in Fig.~\ref{density_dist}, 
i.e., below, in the vicinity, above and far away of the transition
point.
It is remarkable that $S(k)$ exhibits a maximum for all considered
values of the global density at $k_0 \approx 0.72$ (dashed lines in
Fig.~\ref{struc_5}). 
This value correspondence to the characteristic wave length 
$\lambda_0=\frac{2 \pi}{k_0}$ of the density fluctuations 
in the free flow phase.
The steady state structure factor is related to the 
Fourier transform of the real space density-density
correlation function.
The wave length $\lambda_0$ corresponds to a maximum of the
correlation function, i.e., $\lambda_0$ describes the 
most likely distance of two particles in the free flow phase.
For low densities the structure factor is almost independent
of the density and displays a minimum
for small $k$ values indicating the lack of long-range correlations.
Crossing the transition point the smallest mode 
$S(k=\frac{2 \pi}{L})$ increases quickly. 
This suggests that the jammed phase is characterized by 
long-range correlations which decay in the limit $\rho_g \gg \rho_c$ 
algebraically as one can see from the log-log plot 
in Fig.~\ref{struc_5}d.

\begin{figure}
 \begin{minipage}[t]{8.6cm}
 \epsfxsize=8.6cm
 \epsfysize=7.5cm
 \epsffile{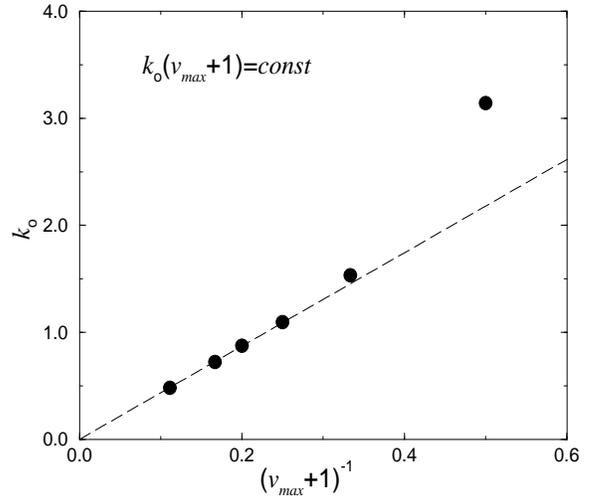} 
 \caption{The characteristic inverse wavelength $k_0$ for
	  the maximum velocities $v_{max}\in\{1,2,3,4,5,8\}$ for $P=\frac{1}{2}$.
 \label{k_max_vmax}}
 \end{minipage}
\end{figure}

We already mentioned that $k_0$ characterizes the 
density fluctuations in the free flow phase.
In Fig.~\ref{k_max_vmax} we plot $k_0$ as a function of
$v_{max}$.
Except for the case $v_{max}=1$ the maximum velocity and 
$k_0$ obey the relation
$k_0 \, (v_{max}\,+\,1) \; = \; const.$
The fact that $v_{max}=1$ violates 
this equation
does not surprise.
It is already known that the physics for $v_{max}\ge 2$
is distinctly different from the case $v_{max}=1$ \cite{SCHRECK_1}.
For instance, for $v_{max}=1$ the fundamental 
diagram is symmetric around its maximum at $\rho_g=0.5$ 
independent of the noise parameter $P$. 
Whereas, the position of the maximum depends on $P$ 
for $v_{max}\ge 2$ and no symmetry occurs.
Another example is that for $v_{max}\ge2$ jams are allowed to 
branch (see Fig.~\ref{space_time_dia}),  
unlike jams for $v_{max}=1$ \cite{NAGEL_2}.
The qualitative different behavior for $v_{max}=1$ 
is caused by a particle-hole symmetry which is lost for 
larger values of the maximum velocity \cite{SCHRECK_1}.

Changing from momentum space to real space the
characteristic wavelength $\lambda_0$ of the density
fluctuations in the free flow phase are given by
\begin{equation}
\lambda_0 \; = \; \frac{2 \pi}{k_{\mbox{o}}}
\; = \; 2 \pi \, \frac{v_{max}\,+\,1}{const}.
\label{eq:wavelength}
\end{equation}
On the other hand the average distance $\bar d$ of the particles is
given by the inverse density $\rho_g^{-1}$.
In the free flow phase the average distance is larger than
the wavelength $\lambda_0$,
$\lambda_0 \; \ll \; \bar d \; = \; 1/ \rho_g$,
i.e.,~the cars can be considered as independent particles.
With increasing density this behavior changes when the 
average distance $\bar d$ is comparable to the
wavelength $\lambda_0 \approx \bar d$.
The critical density is related to the characteristic
wavelength 
\begin{equation}
\rho_c \; = \; \frac{1}{\bar d} \; \approx \;
\frac{1}{\lambda_0} \; \sim \; \frac{1}{v_{max}+1}.
\end{equation}
The fact that the critical density scales with
$v_{max}+1$ is already known for the deterministic
case $P=0$ \cite{NAGEL_1}.

\begin{figure}
 \begin{minipage}[t]{8.6cm}
 \epsfxsize=8.6cm
 \epsfysize=7.5cm
 \epsffile{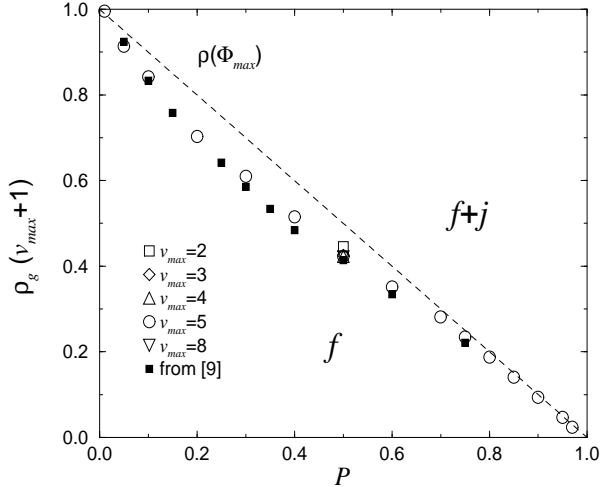} 
 \caption{The phase diagram of the Nagel-Schreckenberg model.
	  Note that in the non-deterministic region $0<P<1$ 
	  the density of the maximum flow exceeds 
	  the density of the transition point.           
 \label{phase_dia}} 
 \end{minipage}
\end{figure}

Up to now we only considered the case $P=\frac{1}{2}$.
The phase diagram in Fig.~\ref{phase_dia} shows the $P$ dependence 
of the transition density $\rho_c$.
{\it f} denotes the free flow phase and
{\it f+j} corresponds to the coexistence region
 where the system separates in the
free flow and jammed phase.
The dashed line displays the $P$ dependence of the maximum
flow obtained from an analysis of the fundamental
diagram \cite{EISEN_1}.
The critical densities $\rho_c$, where the phase transition 
takes place, are lower than 
the density values of the maximum flow.
Measurements of the relaxation time, which is expected to
diverge at a transition point \cite{CSANYI_1},
confirm this result \cite{EISEN_2} (see Fig.~\ref{phase_dia}).
But one has to mention that the determination of the critical
density via relaxation times leads in the coexistence regime
{\it f+j} to unphysical results, in the sense
that the relaxation time becomes negative \cite{EISEN_2}.

\begin{figure}
 \begin{minipage}[t]{8.6cm}
 \epsfxsize=8.6cm
 \epsfysize=7.5cm
 \epsffile{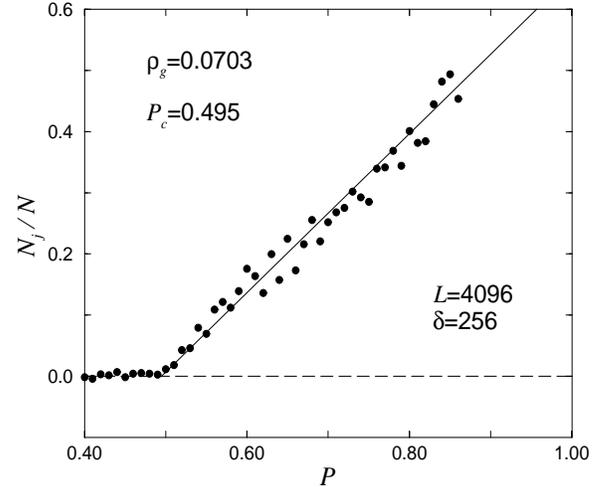} 
 \caption{The number of particles in the high density phase $N_j$ normalized
	  by the total number of particles $N$.
	  $N_j/N$ could serve as an order parameter and the critical
	  exponent is $\beta=1$ (see solid line).
 \label{order_par}} 
 \end{minipage}
\end{figure}

Motivated by real traffic flow we fixed in our
measurements the noise parameter $P$ and varied the 
global density $\rho_g$, i.e.,~we crossed the critical
line in the phase diagram parallel to the vertical
axis (see Fig.~\ref{phase_dia}).
In theoretical investigations however, it is more convenient and revealing 
to consider the crossing of the critical line
parallel to the horizontal axis, i.e., increasing
the noise parameter $P$ by a fixed density.
With growing $P$ density fluctuations with a 
characteristic wavelength $\lambda_0$ occur.
This wavelength $\lambda_0$ grows with increasing $P$
but no phase separation takes place  
until it exceeds at the transition line a
critical wavelength $\lambda_c$. 
Then the system separates into the two coexisting phases
and the amount of particles which belong 
to the jammed phase, $N_j$,
could serve as an order parameter.
As mentioned above the local density distribution displays
two peaks in the heterogeneous phase
and the area under each peak is proportional 
to $N_f$ and $N_j$, respectively. 
Figure~\ref{order_par} displays $N_j$ normalized by the
total number of particles ($N=N_f+N_j$).
Approaching the transition point $p_c$,
$N_j$ arises linear, 
i.e.,~it obeys the equation
$N_j\; \sim \; (P-P_c)^{\beta}$,
where the critical exponent is given by $\beta=1$.
The continuously behavior of $N_j$ and the 
diverging relaxation time \cite{EISEN_2} suggest
that the transition of the traffic model
could describe as a phase transition of second
order.
A detailed analysis of the order parameter and 
the order parameter fluctuations, including a finite-size
analysis, requests further investigations.

\section{Conclusions}

In conclusion we have studied numerically the Nagel-Schreckenberg 
traffic flow model using a local density analysis.
Crossing the critical line of the system
a phase transition takes place from a 
homogeneous regime (free flow phase)
to an inhomogeneous regime which is characterized by a 
coexistence of two phases (free flow traffic and jammed traffic).
The decomposition in the phase coexistence regime
is driven by density fluctuations, provided they exceed
a critical wavelength $\lambda_c$.
The amount of particles in the jammed phase could serve as
an order parameter which arises linear at the transition
point, suggesting that the transition is of
second order.

\end{document}